\documentclass[aps,pre,reprint,showpacs,showkeys,superscriptaddress]{revtex4-1}
\usepackage{graphicx,amsmath,amssymb,bm,natbib,nicefrac}
\makeatletter \def\fnum@figure{\figurename~\thefigure~(color online)} \makeatother
 
\begin{document}
 
\title{Weakly Explosive Percolation in Directed Networks}
\author{Shane Squires}
 \affiliation{Department of Physics, University of Maryland, College Park, MD}
 \affiliation{Institute for Research in Electronics and Applied Physics, University of Maryland, College Park, MD}
\author{Katherine Sytwu}
 \affiliation{Institute for Research in Electronics and Applied Physics, University of Maryland, College Park, MD}
 \affiliation{Rutgers University, New Brunswick, NJ}
\author{Diego Alcala}
 \affiliation{Institute for Research in Electronics and Applied Physics, University of Maryland, College Park, MD}
 \affiliation{University of Northern Colorado, Greeley, CO}
\author{Thomas M.\ Antonsen}
 \affiliation{Department of Physics, University of Maryland, College Park, MD}
 \affiliation{Institute for Research in Electronics and Applied Physics, University of Maryland, College Park, MD}
 \affiliation{Department of Electrical and Computer Engineering, University of Maryland, College Park, MD}
\author{Edward Ott}
 \affiliation{Department of Physics, University of Maryland, College Park, MD}
 \affiliation{Institute for Research in Electronics and Applied Physics, University of Maryland, College Park, MD}
 \affiliation{Department of Electrical and Computer Engineering, University of Maryland, College Park, MD}
\author{Michelle Girvan}
 \affiliation{Department of Physics, University of Maryland, College Park, MD}
 \affiliation{Institute for Research in Electronics and Applied Physics, University of Maryland, College Park, MD}
 \affiliation{Institute for Physical Sciences and Technology, University of Maryland, College Park, MD}
\date{\today}
\pacs{64.60.ah, 64.60.aq, 05.40.-a, 89.75.Da}
 %64.60.ah: percolation in phase transitions 
 %64.60.aq: networks in phase transitions 
 %05.40.-a: statistical physics 
 %89.75.Da: scaling phenomena in complex systems 
\keywords{Explosive percolation, directed networks, complex networks, 
          phase transitions.}
 
\begin{abstract}
Percolation, the formation of a macroscopic connected component, is 
a key feature in the description of complex networks.  The dynamical 
properties of a variety of systems can be understood in terms of 
percolation, including the robustness of power grids and information 
networks, the spreading of epidemics and forest fires, and the 
stability of gene regulatory networks.  Recent studies have shown 
that if network edges are added ``competitively'' in undirected 
networks, the onset of percolation is abrupt or ``explosive.'' The 
unusual qualitative features of this phase transition have been the 
subject of much recent attention.  Here we generalize this previously 
studied network growth process from undirected networks to directed 
networks and use finite-size scaling theory to find several scaling 
exponents.  We find that this process is also characterized by a very 
rapid growth in the giant component, but that this growth is not as 
sudden as in undirected networks. 
\end{abstract}
 
\maketitle 
 
\section{Introduction}
 
A complex network is a collection of nodes, along with a set of edges 
which join pairs of nodes.  In an undirected network, in which each edge 
may be traversed in both directions, the network can be divided into 
distinct connected components.  As edges are successively added to a 
large undirected network, it may transition from a non-percolating phase, 
in which every connected component is microscopic, to a percolating 
phase, in which there is a single ``giant'' component which contains 
a macroscopic fraction of the nodes in the network \cite{newman01}. 
The fraction of nodes in the giant component is the order parameter 
for the percolation phase transition. 
 
The percolation phase transition on undirected networks was independently 
discovered by Solomonoff and Rapoport \cite{solomonoff51} and Erd\H{o}s 
and R\'enyi \cite{erdhos60} and later generalized by other authors 
\cite{newman01,molloy95}.  The network growth process studied by Erd\H{o}s 
and R\'enyi, now the prototypical example of network percolation, may be 
characterized as follows.  The network initially consists of $N \gg 1$ 
nodes and no edges.  Then, on each successive step of the growth process, 
a pair of nodes is selected randomly and an undirected edge is added 
between them.  The size of the largest connected component is recorded and 
the process is repeated.  The percolation phase transition for networks 
grown in this manner is second-order (continuous) in the number of edges 
in the network.  However, recent work by Achlioptas et al.\ demonstrated 
that simple modifications to this growth algorithm can induce surprisingly 
different behavior in the growth of the giant component \cite{achlioptas09}.  
In particular, they found that introducing ``edge competition'' during 
network growth results in ``explosive percolation,'' a delayed, seemingly 
first-order (discontinuous) transition.  
 
The network growth process proposed by Achlioptas et al.\ is designed to 
inhibit the formation of large connected components.  At each step, two 
random candidate edges are considered, with the intention of selecting 
only one of them for addition to the network.  If one of the edges connects 
two nodes in the same component, it is selected automatically because its 
addition would not cause any component to grow.  If the addition of either 
edge would connect two distinct components, the product of the sizes of 
these two components is compared, and only the edge with the smaller 
product is added to the network \footnote{Achlioptas et al.\ present 
another rule in which the sum, rather than the product, of the component 
sizes are used.  Since they found similar results for the two cases, we 
refer only to the product rule.}.  

\begin{figure}[t] 
\centering 
\includegraphics[width=.95\columnwidth]{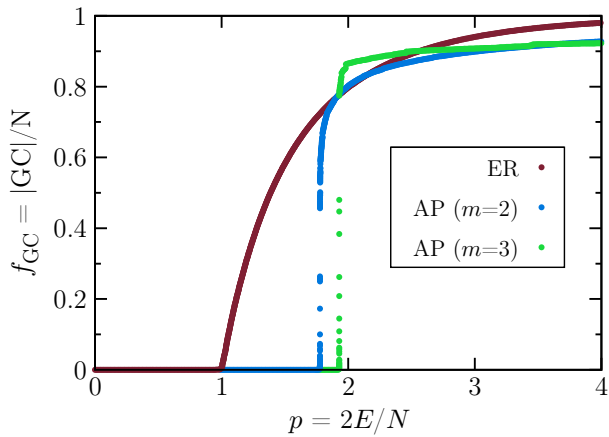}
\caption{The growth of $f_\text{GC}$, the fraction of nodes in the giant 
 component of an undirected network, for three individual networks with 
 $N=2^{23}$.  The growth process is repeated using the Erd\H{o}s-R\'enyi 
 growth process (red or dark gray), the Achlioptas process (blue or medium 
 gray), and a modified Achlioptas process in which three candidate edges, 
 rather than two, are used at each network growth step (green or light gray).}
\label{u-growth}
\end{figure}

Networks grown in this fashion percolate much later than Erd\H{o}s-R\'enyi 
networks; however, when a giant component eventually forms, it grows 
extremely rapidly.  Based on numerical simulations, Achlioptas et al.\ 
conjectured that the phase transition is first-order, but it has now 
been shown that the Achlioptas process actually produces a second-order 
transition \cite{riordan11,grassberger11,da_costa10,nagler11}.  The 
abrupt growth observed in numerical experiments is due to a small but 
positive critical exponent for the growth of the order parameter, along 
with strong finite-size effects which diminish only very slowly as 
$N \to \infty$.  In spite of this, the Achlioptas process continues to 
attract considerable interest because, at network sizes that are typical 
in applications, these finite-size effects give the percolation phase 
transition an ``effectively'' first-order appearance that is qualitatively 
different from that of traditional percolation problems (see Fig.\ 
\ref{u-growth}).  It has also spurred interest in other models which 
exhibit abrupt phase transitions, including Kuramoto \cite{gomez-gardenes11} 
and Ising \cite{angst12} models, as well as other modified percolation 
processes \cite{ziff09,bohman04,chen11,araujo10,schrenk11,boettcher12,
reis12,nagler12,cho13}, many of which are believed to exhibit genuine 
first-order transitions.  

In this paper, we extend the Achlioptas process to \emph{directed} networks.  
In a directed network, each edge can only be traversed in one direction.  
Directed networks are widely used to model gene regulation, food webs, 
neural networks, citation networks, the world-wide web, and other systems.  
However, the existing literature on explosive percolation is exclusively 
focused on undirected networks.  Here, we explore a generalization of the 
Achlioptas process to directed networks and study the scaling properties 
of this process.  We find that competitive edge percolation on directed 
networks shares some of the qualitative features of the Achlioptas process 
on undirected networks, but these features are far less pronounced.  

Because many modified percolation models exhibit unusual phase 
transitions, models with significantly different properties have all 
been labeled explosive in the literature.  For the purposes of this 
paper, the adjective ``explosive'' will refer to the unusual features 
which distinguish the critical behavior of the Achlioptas process from 
both ordinary percolation as well as truly discontinuous models.  These 
features are discussed further in the Results.  We have termed the 
behavior of our directed-network model ``weakly explosive'' because 
it shares some of these qualities, but only to a limited extent.

\section{Methods}
 
\begin{figure}[t] 
\centering 
\includegraphics[width=.95\columnwidth]{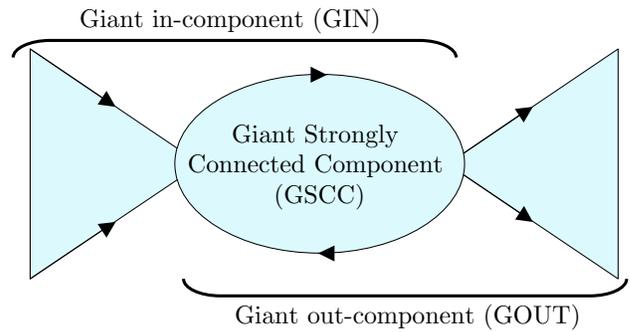}
\caption{An illustration of the ``bow-tie'' structure of the giant 
 component in a directed network above the percolation threshold (see text).}
\label{bowtie}
\end{figure}

In order to define an Achlioptas-like process on directed networks, we 
first need to define connectedness on a directed network.  Although there 
is a single unambiguous definition of a ``connected component'' for 
undirected networks, there are multiple related definitions for directed 
networks \cite{newman01}.  In the algorithms discussed below, we will 
study four different types of structures to which a node may belong.  
In the giant component, these structures are commonly illustrated with 
the well-known ``bow-tie diagram'' (Fig.\ \ref{bowtie}) \cite{broder_et_al.00}. 
First, the in-component of a node $i$, IN($i$), is the set of all nodes 
which have paths to $i$.  Likewise, the out-component of $i$, OUT($i$), 
is the set of all nodes which can be reached on paths from $i$.  Next, 
the strongly connected component of $i$, SCC($i$), is the intersection 
of IN($i$) and OUT($i$).  Finally, we define the full bow-tie, BT($i$), 
to be the union of IN($i$) and OUT($i$) \footnote{Note that the full 
bow-tie of $i$ is not equivalent to the weakly connected component of 
$i$, which is the component to which $i$ would belong if all edges in 
the network were undirected.  This difference can be illustrated by a 
sample directed network of three nodes in which the edges are $1 \to 2$ 
and $3 \to 2$; node $3$ is in the weakly connected component of 1 but 
not BT($1$).  All percolation properties of weakly connected components 
on directed networks are equivalent to those of components on undirected 
networks, so they are not studied here.}.  Each of these structures is 
in some sense analogous to the connected component in undirected networks.  
This comparison extends to the percolation transition in the directed 
Erd\H{o}s-R\'enyi process, in which directed edges are successively 
added between randomly selected, unconnected pairs of nodes.  At the 
critical point, a giant strongly connected component (GSCC), giant 
in-component (GIN), and giant out-component (GOUT) form simultaneously 
\cite{newman01}, comprising the giant bow-tie (GBT).  For convenience 
below, we will use $G$ to denote any one of the parts of the giant 
component of a directed network (GSCC, GIN, GOUT, or GBT), or for the 
giant component (GC) of an undirected network.  See Table \ref{acronyms}
for a list of acronyms.  

Now, we describe a new network growth processes on directed networks. 
We will refer to this process as the directed competition process (DCP) 
to distinguish it from the Achlioptas process (AP), the Erd\H{o}s-R\'enyi 
process (ER), and the directed Erd\H{o}s-R\'enyi process (DER).  It 
also consists of repeatedly choosing two random directed candidate edges 
$i_1 \to j_1$ and $i_2 \to j_2$ from the set of all distinct unoccupied 
edges, then selecting one for addition to the network.  As in the 
Achlioptas process, we automatically select one of the edges if that edge 
is redundant to the connectedness of the network, i.e., if there is already 
a path from $i$ to $j$.  Otherwise, we select the edge for which 
$|\text{IN}(i)| \cdot |\text{OUT}(j)|$ is minimized.  Here, the vertical 
bars denote cardinality, so $|\text{IN}(i)|$ refers to the number of nodes 
in IN($i$).  As in \cite{andrade11,araujo11}, we also consider generalizations 
of both AP and DCP in which $m$ edges (rather than two edges) are chosen for 
consideration at each step in the growth process, and we will discuss results 
for both $m=2$ and $m=3$.  Note that the $m=1$ case of AP corresponds to ER, 
and the $m=1$ case of DCP corresponds to DER.  Finally, in order to emphasize 
that DCP is a generalization of AP, we also note that the two processes are 
identical when applied to an undirected network. 
 
\begin{table}[t] 
\centering 
\includegraphics[width=\columnwidth]{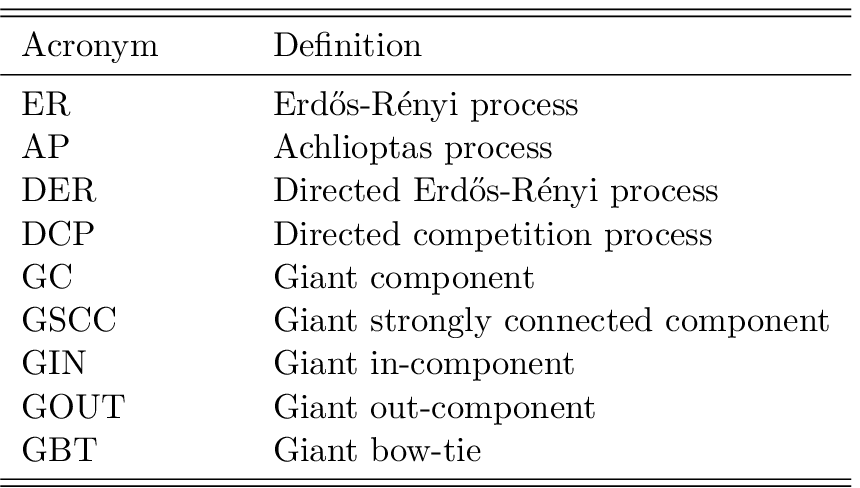}
\caption{Acronyms commonly used in the text.}
\label{acronyms}
\end{table}
 
\begin{figure}[!t] 
\centering 
\includegraphics[width=.95\columnwidth]{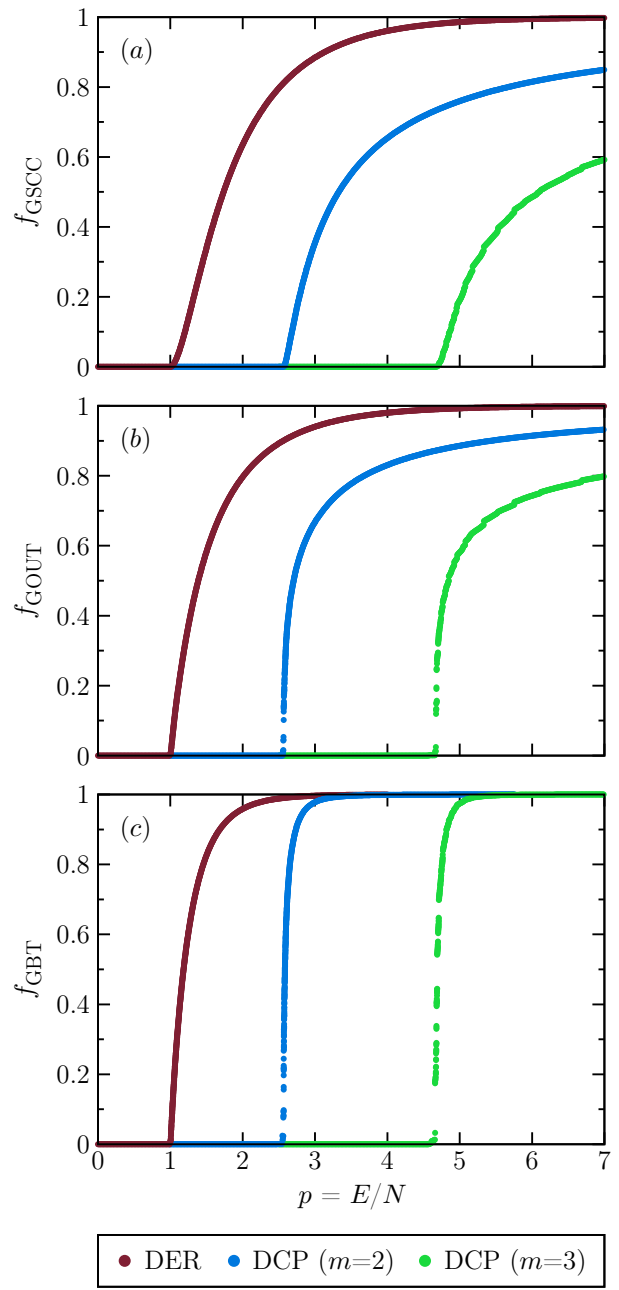}
\caption{The formation of ($a$) the giant strongly connected component, 
 ($b$) the giant out-component, and ($c$) the giant bowtie in a directed 
 network with $N=2^{23}$.  In each panel, the results for the directed 
 Erd\H{o}s-R\'enyi process (red or dark gray) are compared to those for 
 the directed competition process using either $m=2$ (blue or medium gray) 
 or $m=3$ (green or light gray).   Results for GIN are omitted due to 
 symmetry with GOUT.}
\label{d-growth}
\end{figure}
 
The DCP  edge selection rule may also be motivated by noting that it 
minimizes the ``throughput'' which is created by the addition of each 
edge in a way which is analogous to the Achlioptas product rule.  More 
formally, let $P_{ij}$ indicate whether or not there is a path from $i$ 
to $j$, i.e., $P_{ij}=1$ if there is such a path and $P_{ij}=0$ if there 
is not.  The throughput of the network can be defined as $T = \langle P 
\rangle$, where the average is taken over all node pairs $i$ and $j$ 
($i \ne j$).  Well below the percolation threshold, when there are few 
paths from nodes in IN($i$) to nodes in OUT($j$), adding an edge from 
$i$ to $j$ on average increases $T$ by approximately $|\text{IN}(i)| 
\cdot |\text{OUT}(j)|/N^2$.  Similarly, in the Achlioptas process for an 
undirected network, the change in $T$ from the addition of a single 
edge to a network well below the percolation threshold is approximately 
$2 |\text{C}(i)| \cdot |\text{C}(j)|/N^2$, where C($i$) and C($j$) are 
the components to which $i$ and $j$ belong.  Thus, both rules may be 
construed as minimizing $T$ early in the network growth process.  This, 
in turn, leads to an explosive phase transition by creating what has 
been termed a ``powder keg'' \cite{friedman09} of mesoscopic components 
which ``ignites'' at the critical point, when edge competition can no 
longer prevent them from merging. 
 
\begin{table*}[t]
\centering 
\includegraphics[width=\textwidth]{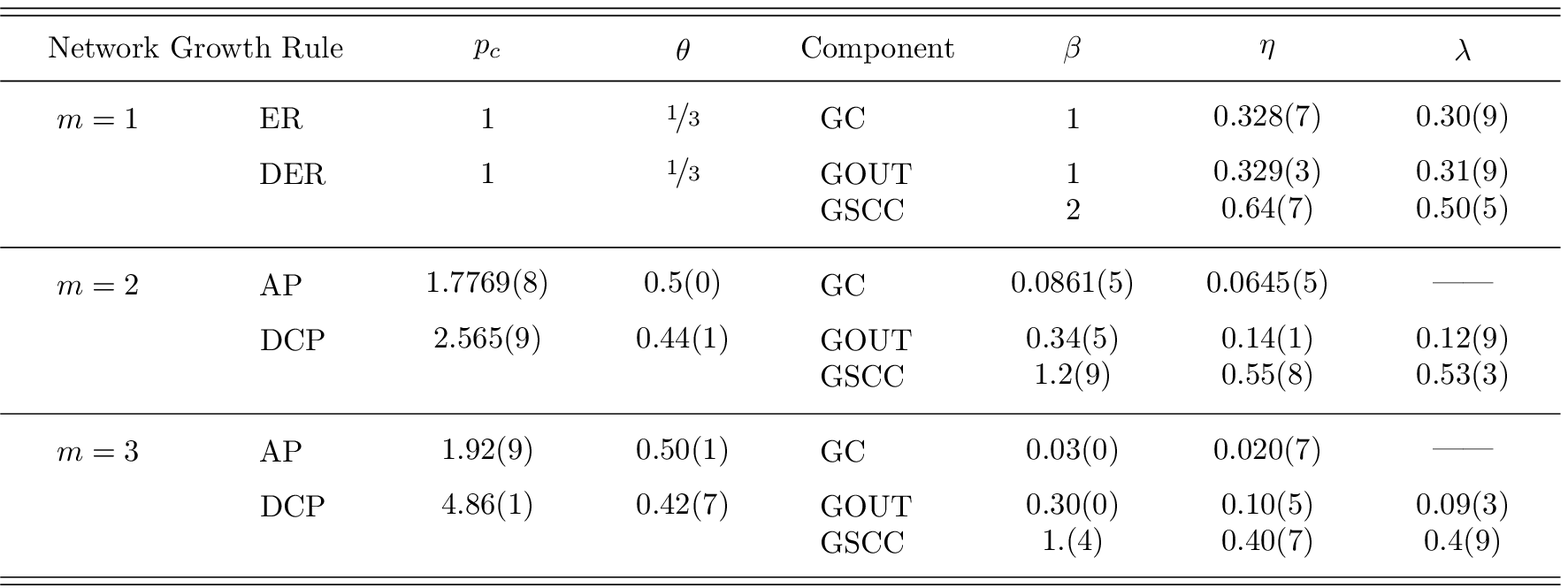}
\caption{Critical exponents for each process (see text).  For ER and DER, 
 $p_c$, $\theta$, and $\beta$ are well-known exact results (see, for example, 
 \cite{erdhos60} and \cite{newman03}).  For AP with $m=2$, we reproduce 
 $p_c$, $\theta$, and $\beta$ from \cite{grassberger11} and $\eta$ from 
 \cite{manna12}; refer to \cite{grassberger11} for additional comments 
 about the interpretation of $\theta$.  All other exponents listed above 
 are derived from our numerical simulations, as described below.  Due to 
 symmetry, results for GIN are identical to those for GOUT, and results 
 for GBT are not listed because, in most cases, they are similar to those 
 for GOUT.}
\label{exptable}
\end{table*}

For the order parameter of each phase transition, we will use the 
normalized size $f_G$ of a giant component, 
\begin{equation}
f_G = \frac{|G|}{N}. 
\end{equation}
We define the GSCC to be the largest strongly connected component in 
the network, the GIN and GOUT to be its in- and out-components, and 
the GBT to be the union of the two \footnote{One consequence of this 
definition is that, while the size of the giant component in undirected 
percolation must increase monotonically as edges are added, this is not 
always true here.  Because GIN, GOUT, and GBT are defined in terms of 
the GSCC, their sizes may decrease if, for example, a strongly connected 
component with a relatively large GIN grows to overtake the GSCC and 
becomes the new GSCC.  These events are relatively rare but do occur.}.  
For the tuning parameter, we will use the average degree of the network, 
$p$.  For undirected networks, $p=2E/N$, whereas for directed networks, 
$p=E/N$ \cite{newman03}.  Note that, for undirected networks, our use of 
$p$ as the tuning parameter differs slightly from the usual convention 
of using $E/N$ as a tuning parameter.  Our use of the average degree is 
motivated by the observation that both undirected and directed Erd\H{o}s-R\'enyi 
networks percolate at the same average degree ($p_c=1$), so $p$ is a 
natural scale for comparison between the directed and undirected cases. 

Computationally, percolation simulations are more time-intensive for 
directed networks than undirected networks.  While only ${\mathcal O}(N)$ 
operations are needed to simulate an entire network growth process in an 
undirected network \cite{newman00}, a na\"ive algorithm for competitive 
edge percolation in a directed network would require at least ${\mathcal O}(N^2)$ 
operations, because there are ${\mathcal O}(N)$ edge additions, between 
each of which several processes with up to ${\mathcal O}(N)$ steps must 
occur.  These processes include checking for a path from $i$ to $j$ for 
each prospective edge $i \to j$, finding IN($i$) and OUT($j$), and decomposing 
the network into strongly connected components \cite{tarjan72}.  In order 
to improve computational performance, we track each part of the giant 
component during the network growth process and use knowledge of the giant 
component to speed up or eliminate the first two processes.  For example, 
if $i$ is in GIN and $j$ is in GOUT, checking for a path from $i$ to $j$ 
is unnecessary because one must exist.  Additionally, we report results 
only for the giant component, not the distribution of other component 
sizes, to avoid the third process.  This results in an algorithm which 
scales approximately as ${\mathcal O}(N^{1.5})$, where most of the time 
is spent in the critical region where more than one macroscopic or 
near-macroscopic component exists.  This improvement enables the 
simulation of networks with significantly larger $N$ than would 
otherwise be feasible.

\section{Results}
 
Plots of the order parameters versus $p$ are shown in Fig.\ \ref{d-growth}
for large-$N$ single-network realizations of the DER and DCP growth 
processes.  When edge competition is present, the emergence of all four 
parts of the giant component are delayed, and the GOUT and GBT display 
sudden growth at the critical point which is qualitatively similar to 
(though less marked than) that of the Achlioptas process (Fig.\ \ref{u-growth}).  
By symmetry, results for GIN are the same as those for GOUT.  In order 
to make quantitative comparisons between DCP and AP, we measure several 
scaling exponents which can be used to characterize the features of 
explosive percolation \cite{grassberger11,da_costa10,radicchi10,nagler11,
manna12}.  In fact, the Achlioptas process is striking precisely because 
these exponents are small (see Table \ref{exptable}), but it is continuous 
because they are nonzero.  

The first such measure is the critical exponent $\beta$, defined by 
\begin{equation}
\langle f \rangle \sim (p-p_c)^\beta 
\end{equation}
as $p \to p_c$ from above, for networks in the thermodynamic limit $N \to 
\infty$.  The average $\langle \cdot \rangle$ is taken over the ensemble 
of grown networks.  Clearly, $\beta>0$ indicates a continuous transition, 
and it has been observed that $0 < \beta \ll 1$ for AP \cite{da_costa10,
grassberger11}.  Next, we report another exponent $\eta$, defined by 
\begin{equation}
\langle \max(\Delta f) \rangle \sim N^{-\eta}, 
\end{equation}
where $\max(\Delta f)$ is the largest jump in $f$ upon the addition of 
a single edge during a network growth process.  In a discontinuous phase 
transition, the maximum jump would approach a nonzero constant as 
$N \to \infty$, corresponding to $\eta=0$, but $\eta$ has also been 
observed to be small and positive for AP \cite{nagler11,manna12}.  

Finally, we introduce a third scaling exponent $\lambda$, defined by 
\begin{equation}
\max_p \left( \text{Var}[f] \right) \sim N^{-\lambda}
\end{equation}
for sufficiently large $N$.  This is motivated by the observation in 
\cite{grassberger11} that, for the Achlioptas process, the maximum 
variance of $f$ initially increases as $N$ grows, then begins to 
decrease very slowly when $N$ is extremely large.  This is related 
to other unusual finite-size effects in AP; see \cite{grassberger11} 
for a thorough discussion.  In a continuous transition, we expect 
that $\text{Var}[f] \to 0$ for all $p$ in the thermodynamic limit, 
so $\lambda>0$.  Moreover, a small value of $\lambda$ indicates that 
for finite $N$, there may be large changes in $f$ near the critical 
point.  

\begin{figure}[!t] 
\centering 
\includegraphics[width=.95\columnwidth]{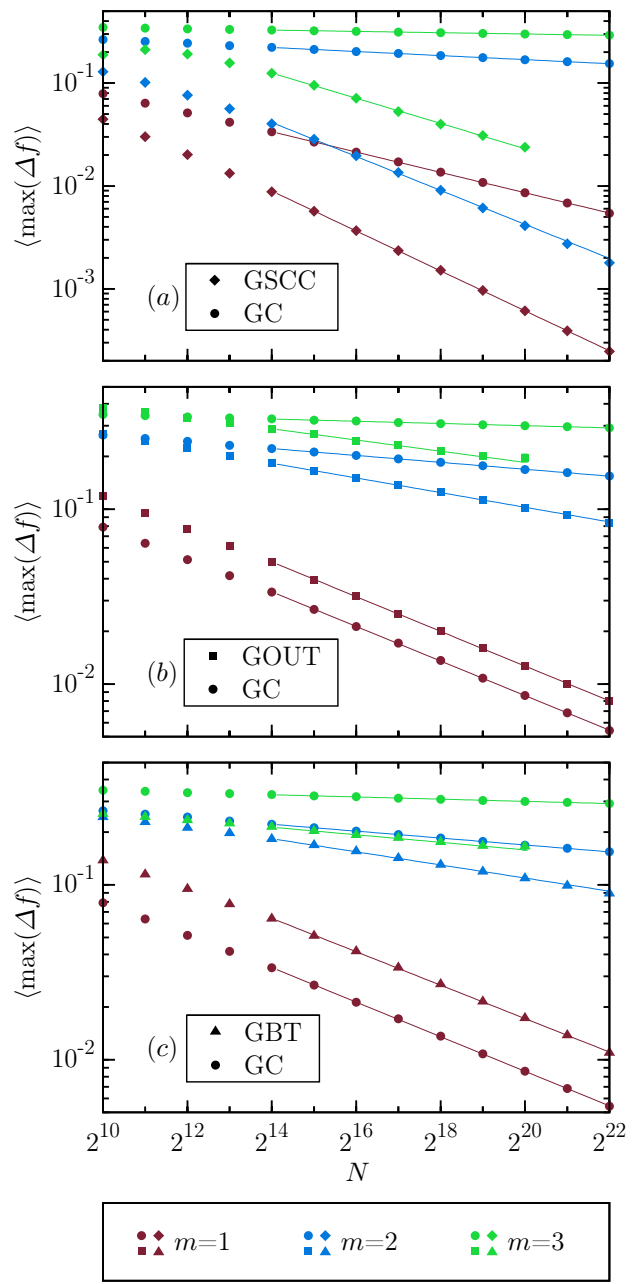}
\caption{Scaling of the maximum jump in the each part of the giant component 
 of directed and undirected networks as a function of $N$.  The results for 
 ($a$) the GSCC (diamonds), ($b$) the GOUT (squares), and ($c$) the GBT 
 (triangles) are compared in each panel to the results for undirected networks 
 (circles).  In addition, ER and DER (red or dark gray) are compared to AP 
 and DCP with $m=2$ (blue or medium gray) and $m=3$ (green or light gray).  
 Lines are power-law fits, whose slopes are given as $\eta$ in Table 
 \ref{exptable}.  Each point is averaged over many network growth trials 
 ($50$ to $10,000$, depending on $m$ and $N$).  Error bars (one standard 
 deviation of the mean) are smaller than the point size for all points.}
\label{maxjump}
\end{figure}

These observations suggest the use of the following descriptions for the 
critical behavior of percolation models for large but finite networks.  
However, we emphasize that these categories are merely useful heuristics 
for describing qualitative behavior, rather than precise definitions. 
\begin{itemize}
\item Discontinuous: Cases in which $\beta=0$, $\eta=0$, and $\lambda=0$.  
\item Explosive: Cases in which $0 < \beta \ll 1$, $0 < \eta \ll 1$, and 
$0< \lambda \ll 1$.  
\item Weakly explosive: Intermediate cases which cannot be clearly designated 
as either ``explosive'' or ``ordinary.''
\item Ordinary: Cases in which $\beta$ is on the order of 1 and $\eta$ and 
$\lambda$ are not small.  (In practice, a natural standard for comparison is 
ER, in which both $\eta$ and $\lambda$ are approximately \nicefrac{1}{3}.) 
\end{itemize}
In order to avoid confusion, we note that our terminology is not directly 
related to the language of \cite{nagler11}, which distinguishes between 
``strongly'' and ``weakly'' discontinuous transitions \footnote{In 
\cite{nagler11}, ``strongly discontinuous'' transitions have a jump discontinuity 
($\beta=0$, $\eta=0$) and ``weakly discontinuous'' transitions are pointwise 
continuous but contain supralinear growth ($0<\beta<1$, $0<\eta<1$).  We use 
the terms explosive and weakly explosive instead, because we are primarily 
interested in behavior which depends on how small the critical exponents are, 
rather than whether or not they are nonzero.}.  

\begin{figure}[t] 
\centering 
\includegraphics[width=.95\columnwidth]{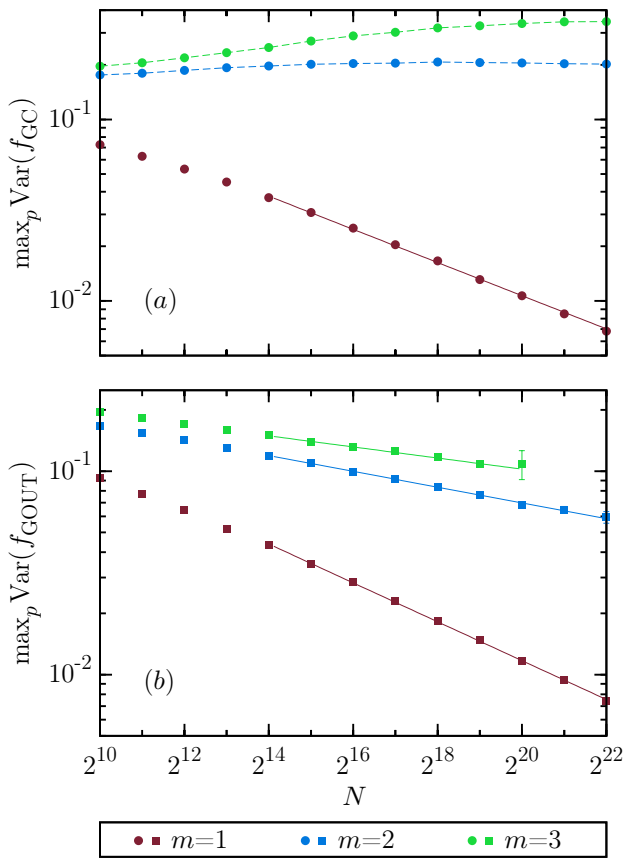}
\caption{Scaling of the largest variance in ($a$) $f_\text{GC}$ (circles)
 and ($b$) $f_\text{GOUT}$ (squares) as a function of $N$, for $m=1$ (red 
 or dark gray), $m=2$ (blue or medium gray), and $m=3$ (green or light gray), 
 using data from the same simulations as in Fig.\ \ref{maxjump}.  Solid 
 lines are power-law fits whose slopes are given as $\lambda$ in Table 
 \ref{exptable}; dashed lines merely connect the data points to guide 
 the eye of the reader.  One unusual feature of AP is that the maximum 
 variance of $f$ increases with $N$, for $N$ not too large (see text), 
 but then eventually decreases; DCP does not share this feature.}
\label{maxvar}
\end{figure}

Now we turn to the numerical estimation of the critical exponents.  Both $\eta$ 
and $\lambda$ may be determined by a straightforward fit to a power law using 
a weighted sum of squares; see Figs.\ \ref{maxjump} and \ref{maxvar}.  The 
critical exponent $\beta$, as well as the critical point $p_c$, are more 
difficult to estimate.  To do this we analyze the finite-size scaling properties 
of the system.  Sufficiently close to the critical point of a continuous phase 
transition, the order parameter $f$ is hypothesized to obey the finite-size 
scaling relation 
\begin{equation}
\langle f \rangle = (p-p_c)^\beta g\left( N^\theta (p-p_c) \right), 
\end{equation}
where $\theta$ determines the scaling of the width of the critical region 
and $g$ is a universal scaling function \cite{newman99,grassberger11}.  This 
may be written in the equivalent form 
\begin{equation}
\langle f \rangle = N^{-\beta \theta} h\left( N^\theta (p-p_c) \right), 
\label{hdef}
\end{equation}
where $h(z) = z^\beta g(z)$ is another universal scaling function.  

\begin{figure}[t]
\centering 
\includegraphics[width=.95\columnwidth]{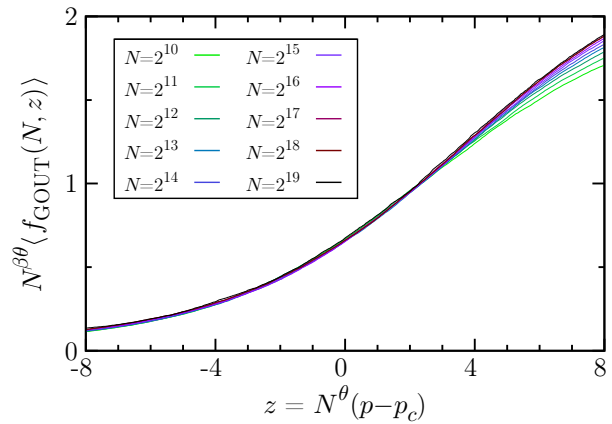}
\caption{Collapse of $\langle f_\text{GOUT} \rangle$ for DCP ($m=2$) 
 using various values of $N$ onto the universal scaling function $h(z)$, 
 according to Eq.\ (\ref{hdef}).  From the bottom curve to the top curve, 
 $N$ increases from $2^{10}$ (light green or light gray) to $N=2^{19}$ 
 (black).  The first eight curves are averaged over 10,000 network growth 
 processes, and the last two are averaged over 5,000 and 2,500 respectively.  
 For these values of $N$, the collapse is excellent up to $z \approx 4$.  
 Similar collapses are used to fit the values of $p_c$, $\theta$, and 
 $\beta$ reported in Table \ref{exptable}.}
\label{collapse}
\end{figure}

Unlike $g(z)$, $h(z)$ is not singular at $z=0$ \cite{newman99,radicchi10}. 
Therefore, Eq.\ (\ref{hdef}) may be interpreted by saying that plots of 
$\langle f \rangle$ versus $z=N^\theta (p-p_c)$ for various values of $N$ 
will all collapse, when appropriately scaled, onto $h(z)$, when $z$ is near 
$0$ (i.e., $p \approx p_c$).  We choose $\beta$, $\theta$, and $p_c$ to 
optimize this data collapse; see Fig.\ \ref{collapse}.  Specifically, we 
choose $\beta$, $\theta$, and $p_c$ to minimize the function 
\begin{equation}
\label{v}
V(\beta, \theta, p_c) = \frac{1}{\Delta z} \int^{\Delta z}_{-\Delta z} 
 \text{Var}_N\left[ N^{\beta \theta} \langle f(z,N) \rangle \right] dz. 
\end{equation}
For further details, see \cite{newman99}.  Since there is no straightforward 
way to estimate the range of validity of Eq.\ (\ref{hdef}), which also depends 
on $N$, we regard $\Delta z$ as an external parameter in Eq.\ (\ref{v}) and 
we choose the value of $\Delta z$ which gives us the smallest minimum in  $V$.  
For all values of $m$ and all components in Table \ref{exptable}, $\Delta z$ 
was between $1$ and $3$.  

The results in Table \ref{exptable} summarize the important features of 
DCP and how they relate to both DER (the analogous non-explosive case) and 
AP (the analogous undirected case).  For the GSCC, $\beta$ and $\eta$ are 
lower in DCP than in DER, but are not small enough to lead to interesting 
behavior; therefore, we will focus on GOUT from here forward.  We see that 
$\beta$ and $\eta$ are significantly smaller in DCP than in DER, but not 
nearly as small as in AP.  This provides quantitative support for our 
characterization of DCP as weakly explosive, in contrast to both the 
explosive behavior of AP and the ordinary behavior of DER.  It is clear 
that DCP belongs somewhere between these two previously-studied regimes.  

Several other features of Table \ref{exptable} are worth noting.  For 
example, in the Achlioptas process, $\beta$ and $\eta$ change quite 
significantly when $m$ is changed from $2$ to $3$, but the corresponding 
changes for DCP are comparatively small.  This suggests again that the 
amount of edge competition has a more pronounced effect on the critical 
behavior of undirected networks than directed networks.  However, the 
opposite is true of the critical point $p_c$, which, for successive values 
of $m$, increases by a much greater factor for directed networks than for 
undirected networks.  If one views the purpose of edge competition as 
\emph{delaying} the formation of a giant component rather than producing 
an explosive transition, then this goal is better achieved by DCP than by AP. 

Finally, in Fig.\ \ref{maxvar}, we see that DCP lacks some of the unusual 
scaling behavior observed for AP in \cite{grassberger11}.  Although the 
values of $\lambda$ for the giant out-component in DCP are smaller than 
those for DER, again indicating weakly explosive behavior, it is nonetheless 
clear that they are positive.  On the other hand, in AP, a much more detailed 
analysis is required to show that $\text{Var}[f]$ eventually approaches 
$0$ for all $p$ as $N \to \infty$ (see \cite{grassberger11}).  Therefore, 
we do not report $\lambda$ for AP, but merely note the qualitative 
differences between AP and DCP.

\section{Discussion}

We have shown that an extension of the Achlioptas process to directed 
networks exhibits critical behavior which is, in many respects, partway 
between classical percolation and explosive percolation, which we have 
termed weakly explosive percolation.  This has several interesting 
ramifications for future research on controlling or modifying percolation 
phase transitions.  One fundamental open question is how general the 
phenomenon of explosive percolation is, and whether the explosiveness of 
a percolation process can be predicted in a relatively straightforward way.  
From the perspective of classical percolation, the primary distinguishing 
features of the Achlioptas network growth process are that it is irreversible 
\cite{da_costa10} and uses nonlocal information \cite{grassberger11}; however, 
there are clearly such processes which are not explosive (see, for example, 
\cite{riordan12}).  The strong explosiveness of the Achlioptas process may be 
contingent on several factors, and the present work suggests that the use 
of undirected networks is one of these factors. 

Another avenue for further research is the possibility of tailoring 
percolation transitions with particular features.  For example, different 
growth rules may create different complex network structures.  In Fig.\ 
\ref{d-growth}, nearly all network nodes have joined the giant bowtie soon 
after the critical point, but this is not true of the giant in- or out-components 
until $p$ is quite large \footnote{In DER, one can show that $f_\text{GBT} = 
f_\text{GOUT}(2-f_\text{GOUT})$ \cite{newman01}, but the behavior of 
$f_\text{GOUT}$ and $f_\text{GBT}$ in Fig.\ \ref{d-growth} is quite different.  
Compared to DER, nodes in DCP are much more likely to be in either GIN or 
GOUT but not in both (that is, not in the GSCC) when $p$ is not too far 
above $p_c$.}.  While it is beyond the scope of this paper to investigate 
this feature, it suggests that there is additional interesting structure 
in networks grown through the directed competition process which cannot 
exist in undirected networks.  More importantly, it may be possible to 
control the critical point and the critical behavior of the giant component 
by using a mix of directed and undirected edges in the network growth process.  
Because the Achlioptas process produces a more explosive transition, but the 
directed competition process delays the onset of criticality for longer, this 
may produce some degree of control for both features.  Along with the above 
results, this suggests that further study of competitive percolation processes 
on directed networks will widen the known repertoire of percolation behavior 
in fascinating ways.

\section{Acknowledgements}
 
This work was supported by ARO grant W911NF-12-1-0101 and is also based 
upon work supported by the National Science Foundation under Grant Number 
PHY-1156454.  Any opinions, conclusions, or recommendations expressed in 
this article are those of the authors, and do not necessarily reflect the 
views of the NSF.

%BibTeX bibliography:
%\bibliography{bib}

%Manual bibliography:
%Merlin.mbs v4.21 2009-07-09.
%

\end{document}